 \newcommand{\be}{\begin{equation}}\newcommand{\ee}{\end{equation}}
\newcommand{\bea}{\begin{eqnarray}}\newcommand{\eea}{\end{eqnarray}}
\newcommand{\nn}{\nonumber}\newcommand{\p}[1]{(\ref{#1})}
\newcommand{\lb}{\label}

\newcommand{\cA}{{\cal A}}
\newcommand{\bcA}{\bar{\cal A}}

\newcommand\T{\mbox{Tr}\;}

\newcommand\Y{{\s Y}}

\newcommand\s{\scriptscriptstyle}
\newcommand\q{\quad}
\newcommand\qq{\qquad}

\newcommand{\pp}{{=\!\!\!|}}
\newcommand{\xp}{x^\pp}
\newcommand{\xm}{x^=}

 \newcommand{\Pp}{\partial_\pp}
\newcommand{\Pm}{\partial_=}

\newcommand{\PY}{\partial_\Y}
\newcommand{\bPY}{\bar\partial_\Y}

\newcommand{\tpi}{\theta^+_i}

\newcommand{\tmi}{\theta^-_i}

\newcommand{\btpi}{\bar{\theta}^{i+}}

\newcommand{\btmi}{\bar{\theta}^{i-}}

\newcommand{\bppt}{\bar{\partial}_{2+}}

\newcommand{\bpmt}{\bar{\partial}_{2-}}

\newcommand{\bpmf}{\bar{\partial}_{4-}}
\newcommand{\bppf}{\bar{\partial}_{4+}}
\newcommand{\ppo}{\partial^1_+}

\newcommand{\pmo}{\partial^1_-}

\newcommand{\pmh}{\partial^3_-}
\newcommand{\pph}{\partial^3_+}

\newcommand{\tpo}{\theta^+_1}
\newcommand{\tpt}{\theta^+_2}
\newcommand{\tmo}{\theta^-_1}
\newcommand{\tmt}{\theta^-_2}
\newcommand{\tmh}{\theta^-_3}

\newcommand{\tmf}{\theta^-_4}
\newcommand{\tpf}{\theta^+_4}
\newcommand{\btpo}{\bar\theta^{1+}}
\newcommand{\btmo}{\bar\theta^{1-}}
\newcommand{\btpt}{\bar\theta^{2+}}
\newcommand{\btmt}{\bar\theta^{2-}}
\newcommand{\btph}{\bar\theta^{3+}}
\newcommand{\btmh}{\bar\theta^{3-}}
\newcommand{\btpf}{\bar\theta^{4+}}
\newcommand{\btmf}{\bar\theta^{4-}}
\newcommand{\Dpo}{D_+^1}
\newcommand{\Dmo}{D_-^1}
\newcommand{\Dpt}{D_+^2}

\newcommand{\bDpo}{\bar{D}_{1+}}

\newcommand{\bDpt}{\bar{D}_{2+}}
\newcommand{\bDmt}{\bar{D}_{2-}}
\newcommand{\bDph}{\bar{D}_{3+}}

\newcommand{\bDpf}{\bar{D}_{4+}}
\newcommand{\bDmf}{\bar{D}_{4-}}
\newcommand{\Dot}{D^1_2}
\newcommand{\Dto}{D^2_1}
\newcommand{\Vot}{V^1_2}
\newcommand{\Vto}{V^2_1}
\newcommand{\Doh}{D^1_3}

\newcommand{\Dtf}{D^2_4}

\newcommand{\Dhf}{D^3_4}
\newcommand{\Dfh}{D^4_3}

\newcommand{\Vfh}{V^4_3}

\newcommand{\N}{\nabla}
\newcommand{\bN}{\bar\nabla}

\documentstyle[12pt]{article}

\begin{document}
\begin{center}
{\large\bf
SHORT HARMONIC SUPERFIELDS AND LIGHT-CONE GAUGE IN  SUPER-YANG-MILLS
EQUATIONS \footnote{To be published in Proceedings of the conference
"Quantization, gauge theories and strings" dedicated to the memory of
E.S. Fradkin, Moscow, June 5-10, 2000.}  }
\vspace{0.5cm}

{\bf B.M. Zupnik}\\

{\it Bogoliuibov Laboratory of Theoretical Physics, Joint Institute
for Nuclear Research, Dubna, Moscow Region, Russia; e-mail:
zupnik@thsun1.jinr.ru}

\end{center}
\begin{abstract}
We analyze the superfield equations of the 4-dimensional $N{=}2$ and
$N{=}4$ SYM-theories using light-cone gauge conditions and the
harmonic-superspace approach. The  harmonic superfield equations of
motion are drastically simplified  in this gauge, in particular, the
basic harmonic-superfield matrices  and the corresponding harmonic
analytic gauge connections become nilpotent on-shell.
\end{abstract}
\renewcommand{\thefootnote}{\arabic{footnote}}
\setcounter{footnote}0
\setcounter{equation}0
\section{Introduction}

The $D=4,~N=2$  harmonic-superspace (HSS)  has been introduced
first for the solution of the  off-shell superfield constraints
\cite{GIK1}. The superfield action and equations of motion of the
$N=2$ super-Yang-Mills (SYM) theory have been also constructed
in this superspace \cite{GIOS,Zu2}.
In the standard harmonic formulation of this theory, the basic harmonic
connection satisfies the conditions of the Grassmann (G-) analyticity,
and the 2-nd one ( via the zero-curvature condition) appears to be a
nonlinear function of the basic connection. The $N=2$ equation of motion
is linearly dependent on the 2-nd harmonic connection, but it is the
nonlinear equation for the basic connection. It has been shown in
Ref.\cite{Zu4} that one can alternatively choose the 2-nd harmonic
connection as a  basic superfield, so that the dynamical G-analyticity
condition (or the Grassmann-harmonic zero-curvature condition) for the
first connection becomes a new equation of motion. In the harmonic approach
to the $D=4,\;N=3$ SYM-theory \cite{GIKOS}, the $SU(3)/U(1)\times U(1)$
harmonics have been used for the covariant reduction of the spinor
coordinates and derivatives and for the off-shell description of the
SYM-theory in terms of the corresponding G-analytic superfields.
Moreover, it was shown that the $N=3$ SYM-constraints in the ordinary
superspace \cite{So} can been transformed to the zero-curvature equations
for the analytic harmonic gauge connections (see, also the alternative
formalism \cite{Wi}).
Harmonic superspaces of the $D=4, N=4$ supersymmetry have been considered
in Refs. \cite{Ba,HH,AFSZ}.

 The short on-shell harmonic superfields
describing the Abelian $N=2, 3$ and 4 SYM-multiplets satisfy the
constraints of chirality or different types of harmonic and Grassmann
analyticities.  It will be shown that the short harmonic superfields
and the nonlinear harmonic equations can be used to describe the classical
solutions of the non-Abelian SYM-equations for different $N$.

We shall analyze the classical solutions of the harmonic-superfield
equations  using the convenient light-cone gauge conditions for superfield
connections (see, e.g. \cite{BLN,DL,GS}). As it has been shown in Refs.
\cite{NZ} these gauge conditions yield the nilpotent superfield matrices
in the  bridge representation of the $N=3$ SYM-theory and simplify
drastically the harmonic-superfield equations. We shall consider
analogous nilpotent gauge conditions for the $N=2$ and $N=4$ theories
and the corresponding simplifications of the harmonic SYM-equations.

\setcounter{equation}0
\section{\lb{B}Solving  $D=4,~N=2$ SYM
 equations in harmonic superspace}

The covariant coordinates of the $D=4,~N=2$ superspace are
\be
z^M=(x^{\alpha\dot\alpha} ,\theta^\alpha_i ,\bar\theta^{i\dot\alpha} )~,
\lb{A2b}
\ee
where $\alpha,~\dot\alpha$ are the $SL(2,C)$ indices
and $i=1, 2$  are indices of the fundamental
representations of the  group $SU(2)$.

We shall study solutions of the  SYM-equations using the non-covariant
notation
\bea
&&\xp\equiv x^{1\dot{1}} =t+x^3~,\q \xm\equiv x^{2\dot{2}}=t-x^3~,\q
y\equiv x^{1\dot{2}}=x^1+ix^2~,
\nn\\
&&\bar{y}\equiv x^{2\dot{1}} =x^1-ix^2~,\q(\tpi,~\tmi)\equiv
\theta^\alpha_i~, \q (\btpi,~\btmi)\equiv  \bar\theta^{i\dot\alpha}
~.\lb{A2}
\eea
suitable when the Lorenz symmetry is reduced to $SO(1,1)$.
The general $N=2$ superspace has the odd dimension (4,4) in this
notation.

Let us  define the  gauge connections $A(z)$ and the corresponding
covariant derivatives $\nabla$ in the $(4|4,4)$-dimensional superspace,
then the $D=4,~N=2$ SYM-constraints \cite{GSW} have the following
reduced-symmetry form:
\bea
&&\{\nabla^k_+,\nabla^l_+\}=0~,\q \{\bar{\nabla}_{k+},\bar{\nabla}_{l+}\}
=0~,
\q \{\nabla^k_+,\bar{\nabla}_{l+}\}=2i\delta^k_l\nabla_\pp~,\lb{A8}\\
&&\{\nabla^k_+,\nabla^l_-\}=\varepsilon^{kl}\bar{W}
~,\q\{\nabla^k_+,\bar{\nabla}_{l-}\}=2i\delta^k_l \nabla_y~,\lb{A9}\\
&&\{\nabla^k_-,\bar{\nabla}_{l+}\}=2i\delta^k_l\bar{\nabla}_y~,\q
\{\bar{\nabla}_{k+},\bar{\nabla}_{l-}\}=\varepsilon_{kl}W~,\lb{A10}\\
&&\{\nabla^k_-,\nabla^l_-\}=0~,\q \{\bar{\nabla}_{k-},\bar{\nabla}_{l-}\}
=0~,
\q \{\nabla^k_-,\bar{\nabla}_{l-}\}=2i\delta^k_l\nabla_=~,\lb{A11}
\eea
where $W$ and $\bar{W}$ are the gauge-covariant
superfield strengthes constructed from the gauge connections. In particular,
this reduced form of the 4D constraints is convenient for the study of
dimensional reduction.

The Bianchi identities yield the following relations for
 the superfield strengthes
\bea
&& \bar\nabla_{i\pm}W=0~,\q \nabla^i_\pm \bar{W}=0
~,\\
&& \nabla^{(i}_+\nabla^{k)}_-W=\bar\nabla^{(i}_-\bar\nabla^{k)}_+\bar{W}~.
\lb{A12}
\eea

Let us analyze first Eqs.\p{A8} together  with the relations
\be
[\nabla^k_+,\nabla_\pp]=[\bar\nabla_{k+},\nabla_\pp]=0~.
\ee

These equations for the positive-helicity connections have the
 pure gauge solutions only, so the simplest light-cone gauge condition  can
be taken in the following form:
\be
A^k_+=0~,\q \bar{A}_{k+}=0~,\q A_\pp =0~.\lb{A14}
\ee

Let us consider the covariantly chiral superfield strength
in the gauge \p{A14}
\be
W={1\over2}\bar{D}^k_+\bar{A}_{k-}~.
\ee
The nonlinear superfield equation of motion for the $N=2$  non-Abelian
SYM-theory has  dimension $l=-2$
\be
D^{(i}_+\nabla^{k)}_-W=0~.\lb{cbequ}
\ee
In distinction with the case $N=3$ \cite{So}, this dynamical equation is
independent of constraints (\ref{A8}-\ref{A11}).

The Lorenz-covariant  $SU(2)/U(1)$ harmonic superspace was introduced
in Ref.\cite{GIK1} for the off-shell description of the $4D\; N=2$
SYM-theory, supergravity and hypermultiplets.

Now we shall  study  this  harmonic superspace in
another ( reduced-symmetry) representation which allows us
to consider the non-covariant gauges and the dimensional reduction.

The $SU(2)/U(1)$ harmonics \cite{GIK1} parametrize the sphere $S^2$.
They form an $SU(2)$ matrix $u^I_i$ and are defined modulo $U(1)$
\be
u^1_i=u_{2i}\equiv u^+_i\;,\q u^2_i=-u_{1i}\equiv u^-_i\;.\lb{F1}
\ee
where $i=1, 2$ is the index of the fundamental representation of $SU(2)$.
Note that we use the non-standard notation of the $U(1)$-charges in
comparison to Ref.\cite{GIK1} since indices $\pm$ are reserved for the
$SO(1,1)$ weights in this paper.

The $SU(2)$-invariant harmonic derivatives act on the harmonics
$\partial^I_J u^K_i =\delta^K_J u^I_i$.

We can define a non-covariant notation for coordinates in the analytic
harmonic superspace $H(4,2|2,2)$
\bea
&&\zeta=(X^\pp,~X^=,~Y,~\bar{Y}~| \theta^\pm_2\;,
\bar\theta^{1\pm})\;,\qq
X^\pp=\xp +i(\tpt\btpt -\tpo\btpo)\;,\nn\\
&&X^= =\xm +
i(\tmt\btmt -\tmo\btmo)\;,\q Y=y+i(\tpt\btmt -\tpo\btmo)\;,\nn\\
&&\bar{Y}=\bar{y}+i(\tmt\btpt -\tmo\btpo)\;,\q
\theta^\pm_I=\theta^\pm_k u^k_I~,\q\bar\theta^{I\pm}=
\bar\theta^{\pm k}u_k^I~.
\lb{F6}
\eea

The special $SU(2)$-covariant conjugation of harmonics preserves the
$U(1)$-charges
\be
\widetilde{u^1_i}=u^i_2\;,\q \widetilde{u^2_i}=-u_1^i\;,\q
\widetilde{u_2^i}=-u^1_i~,\q \widetilde{u_1^i}=u^2_i~.\lb{conj1}
\ee
On the harmonic derivatives of an arbitrary harmonic function $f(u)$ this
conjugation acts as follows
\be
\widetilde{\partial^1_2 f}=\partial^1_2\widetilde{f}\;,\q
\widetilde{\partial_1^2 f}=\partial^2_1\widetilde{f}\;.
\lb{conj2}
\ee

The tilde-conjugation of the odd analytic coordinates has the following
form:
\be
\theta^\pm_1~\rightarrow~\bar\theta^{2\pm}~,\q \bar\theta^{2\pm}
~\rightarrow~-\theta^\pm_1~,\q
\theta^\pm_2~\rightarrow~-\bar\theta^{1\pm}~,\q
\bar\theta^{1\pm}~\rightarrow~\theta^\pm_2~.\lb{conj3}
\ee
Coordinates $X^\pp$ and $ X^=$ are real and $\widetilde{Y}=
\bar{Y}$.

The corresponding CR-structure involves the derivatives
\be
D^1_\pm,\;\bar{D}_{2\pm},~\Dot
\lb{F7}
\ee
which have the following explicit form in these coordinates:
\bea
&& D^1_\pm=\partial^1_\pm~,\q
 \bar{D}_{2\pm}=\bar\partial_{2\pm} \;,\lb{F8}\\
&& \Dot =\partial^1_2
+2i\tpt\btpo\Pp+2i\tpt\btmo\PY+2i\tmt\btpo\bPY
+2i\tmt\btmo\Pm \nn\\
&&-\tpt\ppo-\tmt\pmo+\btpo\bppt+\btmo\bpmt~,
\lb{F9}
\eea
where $\Pp =\partial/\partial X^\pp,~\Pm =\partial/
\partial X^=,~\PY=\partial/\partial Y$ and $\bPY=
\partial/\partial\bar{Y}$.

It is crucial that we start from  the light-cone  gauge conditions
 for the $N=2$ SYM-connections which break  $SL(2,C)$, but preserve
 $SU(2)$. Consider the harmonic transform of the  covariant Grassmann
derivatives  via the projections on the $SU(2)$-harmonics. As result we
get so called harmonized Grassmann covariant derivatives
\bea
&&\nabla^I_+\equiv u_i^I D^i_+=D^I_+\;,\q
\bar\nabla_{I+}\equiv u^i_I\bar{D}_{i+}=\bar{D}_{I+}\;,\q
\{D^I_+,\bar{D}_{K+}\}=2i\delta^I_K\Pp\;,\lb{D1b}\\
&&\nabla^I_-\equiv u_i^I\nabla^i_-=D^I_- +\cA^I_-\;,\q
\bar\nabla_{I-}\equiv u^i_I\nabla_{i-}=\bar{D}_{I-} +\bcA_{I-}\;.
\lb{D1}
\eea

The $SU(2)$-harmonic projections of the superfield constraints
(\ref{A9}-\ref{A11}) can be derived from the basic set of the
$N=2$ zero-curvature (or G-integrability) conditions for two harmonized
 Grassmann connections:
\bea
&&\Dpo\cA^1_-=\bDpt\cA^1_-=\Dpo\bcA_{2-}=\bDpt\bcA_{2-}=0~,\lb{D2}\\
&&\Dmo\cA^1_- +(\cA^1_-)^2=0\;,\q \bDmt\bcA_{2-}+(\bcA_{2-})^2=0\;,\nn\\
&&\Dmo\bcA_{2-}+\bDmt\cA^1_- +\{\cA^1_-,\bcA_{2-}\}=0\;.\lb{D3}
    \eea
All projections of the SYM-constraints can be obtained   acting
by the harmonic  derivatives $D^2_1$ on this basic set of conditions.

The G-integrability equations \p{D3} have a very simple general
solution, namely
\be
\cA^1_-(v)=e^{-v}\Dmo e^v\;,\q \bcA_{2-}(v)=e^{-v}\bDmt e^v\;,\lb{D4}
\ee
where {\it the bridge } $v$ is a
superfield matrix satisfying the additional  constraint
\be
(\Dpo,\;\bDpt) v=0\;,\lb{D5}
\ee
which is compatible with the light-cone representation \p{D1b}. Thus,
$v$ does not depend on $\tpo$ and $\btpt$ in  analytic coordinates
\p{F6}.

Consider the gauge transformations of  bridge $v$
\be
e^v\;\Rightarrow\;e^\lambda e^v e^\tau\;,\lb{D11}
\ee
where $\lambda\in H(4,6|2,2)$ is a G-analytic matrix parameter,
and  constrained parameter $\tau$  does not depend on
harmonics.
Matrix $e^v$  can be interpreted as a map  of  gauge superfields
$ A^k_\pm, \bar{A}_{k\pm}$ defined in the central
basis (CB) to  those in the analytic basis (AB).

The  SYM-constraints for $v$ due to Eqs. \p{D4}
are reduced to the following harmonic differential conditions for
the basic Grassmann connections:
\be
D^1_2\left(\cA^1_-(v),\;\bcA_{2-}(v)\right)=0\;.
\lb{hanal}
\ee
Note that these {\it H-analyticity} relations are trivial for
Grassmann connections in the CB-superfield representation, but they
become the nontrivial differential equations for  bridge $v$.
Equations (\ref{hanal}) are completely equivalent to
the  G-integrabity equations \p{D3}, and they can be treated as a new
representation of the SYM-constraints.

It is not difficult to built all Grassmann CB-connections  in  terms of
basic ones
\bea
&&\cA_-^2(v)=D^2_1\cA_-^1(v)\;,\nn\\
&&\bcA_{1-}(v)=-D^2_1\bcA_{2-}(v)\;.\lb{rcb1}
\eea

The  non-Abelian CB-superfield strengths are
\bea
&&W=\bDpo\bcA_{2-}(v)=-\bDpt\bcA_{1-}(v)\;,
\nn\\
&&\bar{W}=-\Dpt\cA_-^1(v)=\Dpo\cA_-^2(v)~.
\lb{rcb2}
\eea

Now we shall determine the explicit form of  bridge $v$. Using
the off-shell $(4,4)$-analytic $\lambda$-transformations
$\delta v=\lambda +{1\over2}[\lambda,v]+\ldots$ one can
choose the following non-supersymmetric nilpotent  gauge condition for $v$
\bea
&&v=\tmo b^1+ \btmt \bar{b}_2 +\tmo\btmt d^1_2\;,\q v^2=\tmo\btmt
[\bar{b}_2,b^1]\;,\q v^3=0~,\lb{D12c}\\
&&e^{-v}= I - v+{1\over2}v^2=I-\tmo b^1- \btmt \bar{b}_2+\tmo\btmt
({1\over2}[\bar{b}_2,b^1]-d^1_2 ) \;,
\lb{D12b}
\eea
where  fermionic   matrices $b^1, \bar{b}_2 $ and  bosonic matrix
$d^1_2 $ are analytic functions of  coordinates $\zeta$ \p{F6}.

Note that the analogous nilpotent gauge for the $N=3$ harmonic bridge
has been introduced in Ref.\cite{NZ}.  This gauge for $v$ combines the
actions of both the analytic gauge group and the CB-gauge group.

We  use the  harmonic tilde-conjugation  (\ref{conj1},
\ref{conj3}) to describe the reality conditions for the gauge superfields
in HSS. For instance, the Hermitian conjugation $\dagger$ of the
superfield matrices includes transposition and this conjugation.
The  conditions for bridge $v$ in the gauge group  $SU(n)$
 are $v^\dagger=-v$ and $ \T v=0 $, so that
 matrices  $b^1, \bar{b}_2 $ and $d^1_2 $ have
the following properties in the group $SU(n)$:
\bea
&& \T b^1=0\;,\q \T \bar{b}_2=0\;,\q \T d^1_2=0\;
,\lb{D13}\\
&&(b^1)^\dagger=\bar{b}_2\;,\q (\bar{b}_2)^\dagger=-b^1~,
\q (d^1_2)^\dagger=d^1_2~.
\lb{D14}
\eea
Note that the last property is connected with
relation $(\tmo\btmt)^\dagger=-\tmo\btmt$.

It is useful to consider the explicit parametrization of the Grassmann
connection $\cA^1_-(v)$ and $\bcA_{3-}(v)$ in terms of  basic analytic
matrices \p{D12c}
\bea
&&\cA^1_-(v)\equiv e^{-v}\Dmo e^v=b^1-\tmo (b^1)^2+\btmt f^1_2
+\tmo\btmt[b^1,f^1_2]\,, \lb{K9}\\
&&\bcA_{2-}\equiv e^{-v}\bDmt e^v=\bar{b}_2 +\tmo \bar{f}^1_2
-\btmt (\bar{b}_2)^2
+\tmo\btmt[\bar{f}^1_2,\bar{b}_2]~,
\eea
where the following auxiliary superfields are introduced:
\be
f^1_2=d^1_2-{1\over2}\{b^1,\bar{b}_2\}~,\q\bar{f}^1_2=d^1_2+{1\over2}
\{b^1,\bar{b}_2\}~.\lb{K9b}
\ee

Equation $\Dot \cA^1_-(v)=0$ generates the following
independent relations for the (2,2)-analytic matrices:
\bea
&& \Dot b^1=-\tmt (b^1)^2-\btmo f^1_2\;,\lb{D17}\\
&&\Dot f^1_2=\tmt[f^1_2,b^1]~.
\lb{D19}
\eea
Equation $\Dot \bcA_{2-}(v)=0$ gives the conjugated
relations.

It is easily to show that these equations yield the subsidiary
conditions for the coefficient functions
\be
\Dot\Dot b^1=2\tmt\btmo[f^1_2,b^1]\;,\q (\Dot)^3 b^1
=0\;,\q (\Dot)^2f^1_2=0~.\lb{lin2}
\ee

In order to understand more deeply the geometric structure of our
harmonic-superspace solutions it is useful to represent them in
the analytic basis. Remember that the following covariant Grassmann
derivatives are flat in the analytic representation of the gauge group
 before the gauge fixing:
\be
e^v\nabla^1_\pm e^{-v}=D^1_\pm\;,\q
e^v\bar{\nabla}_{2\pm} e^{-v}=\bar{D}_{2\pm}\;.
\lb{H0}
\ee

The harmonic transform of the covariant derivatives via  matrix
$e^v$ \p{D4} determines in AB the on-shell harmonic connections as a
function of $v$
\be
\nabla^I_K\equiv e^v D^I_K e^{-v}=D^I_K+V^I_K(v)~,\q I\neq K~.
\ee

In the off-shell $N=2$ formalism \cite{GIK1}, the connection
$V^1_2$ is the basic G-analytic prepotential by construction. The nonlinear
solution for the 2-nd harmonic connection $V^2_1(V^1_2)$ can be found
explicitly \cite{Zu2}. The Grassmann connections can be constructed via
this connection
\be
a^2_\pm=-D^1_\pm V^2_1~,\q \bar{a}_{1\pm}=\bar{D}_{2\pm}V^2_1~.\lb{Aconn}
\ee

In this analytic representation, the  equation of motion of the
SYM-theory has the following form:
\be
\Dpo\Dmo\bDpt\bDmt V^2_1(V^1_2)=0~.\lb{eq1}
\ee
Stress that this dynamical equation is nonlinear, while the Grassmann
analyticity of $\Vot$ is pure kinematic (off-shell).

The alternative non-analytic representation of  harmonic-superspace
equations for the $N=2$ SYM-theory has been proposed in Ref.\cite{Zu4}.
The basic superfield variable of this dual representation is
non-analytic connection $V^2_1$, then  equation \p{eq1} becomes the simple
linear constraint. The basic dynamical equation of this
representation has the form of the Grassmann-harmonic zero-curvature
equation, so the structure of the $N=2$ SYM-equation become similar
to the structure of the $=3$ equation in the HSS-approach.

Thus,  the bridge matrix  can be interpreted as the 3-rd possible form
of the basic superfield variable which admits the most simple
  gauge conditions \p{D12c}.

The analytic SYM-equations (\ref{D17},\ref{D19}) are equivalent to
 the following relation :
\be
V^1_2\equiv e^v\Dot e^{-v}=\tmt b^1-\btmo\bar{b}_2\;,\lb{dyn}
\ee
which determines the nilpotent analytic connection in our representation.
One can check straightforwardly the properties of this representation
\be
(V^1_2)^3=0~,\q\Dot V^1_2=\tmt\btmo\{b^1,\bar{b}_2\}~,\q (\Dot)^2 v^1_2=0~.
\ee

The bridge representation of the harmonic connection $V^2_1$ is
\be
 V^2_1(v)=-\tmo \Dto b^1-\btmt \Dto\bar{b}_2
+\tmo\btmt\left(-\Dto d^1_2+{1\over2}\{b^1,\Dto\bar{b}_2\}
-{1\over2}\{\Dto b^1,\bar{b}_2\}\right)~.\lb{nonan4}
\ee

 Equation \p{cbequ}  transforms to the following
simple analytic equation in the bridge representation:
\bea
&&\bDpt\bDmt\Dpo\Dmo V^2_1(v)=2i\bDmt\Dmo\left(e^v\Pp e^{-v}\right)\nn\\
&&=2i\left(-\Pp d^1_2+{1\over2}\{b^1,\Pp\bar{b}_2\}
-{1\over2}\{\Pp b^1,\bar{b}_2\}\right)=0~,\lb{abequ}
\eea
where $\Pp=\partial/\partial X^\pp$. This equation can be interpreted as
a solvable linear equation for superfield $d^1_2$, and one should analyze
it together with the harmonic equations for the basic analytic matrices.

Note that the simple Ansatz $\Pp d^1_2=\Pp b^1=\Pp \bar{b}_2=0$
yields the partial solution of this equation. The corresponding solutions
of the $N=2$ SYM-equations can be obtained from the pure harmonic
equations (\ref{D17},\ref{D19}).

It is important that these differential harmonic equations contain the
nilpotent elements $\tmt$ or $\btmo$ in the nonlinear parts, so the
simplest iteration procedure for finding their solutions  can be obtained
via a partial Grassmann decomposition. This  decomposition
generates the finite set of solvable linear iterative equations.

\setcounter{equation}0
\section{Harmonic-superspace formulation of   $N=4$ SYM
equations}

The harmonic superspaces for the $N=4$ SYM-theory have been discussed
in Refs.\cite{Ba,HH,AFSZ}. It has been shown that the G- and H-analytic
Abelian on-shell superfield strength  lives in the harmonic
superspace with (4+4) Grassmann coordinates. We shall use the analogy with
the HSS description of the $N=3$ SYM-equations \cite{NZ} and consider
the gauge invariance and geometric structure of the superfield $N=4$
equations. Stress that we do not know the off-shell
superfield structure of the $N=4$ SYM-theory in contrast to the
$N=3$ case \cite{GIKOS}, although these theories have the same  physical
component fields.

In order to study the light-cone SYM-conditions, one can use
the non-covariant representation of the $D=4,~N=4$ Grassmann coordinates
 $\theta^\pm_i~, \bar\theta^{i\pm}$
where $i=1, 2, 3, 4$  are indices of the fundamental
representations of  $SU(4)$.

 The $D=4,~N=4$ SYM-constraints \cite{So}
have the following reduced-symmetry form:
\bea
&&\{\nabla^k_+,\nabla^l_+\}=0~,\q \{\bar{\nabla}_{k+},\bar{\nabla}_{l+}\}
=0~,
\q \{\nabla^k_+,\bar{\nabla}_{l+}\}=2i\delta^k_l\nabla_\pp~,\lb{a8}\\
&&\{\nabla^k_+,\nabla^l_-\}=W^{kl}
~,\q\{\nabla^k_+,\bar{\nabla}_{l-}\}=2i\delta^k_l \nabla_y~,\lb{a9}\\
&&\{\nabla^k_-,\bar{\nabla}_{l+}\}=2i\delta^k_l\bar{\nabla}_y~,\q
\{\bar{\nabla}_{k+},\bar{\nabla}_{l-}\}=W_{kl}~,\lb{a10}\\
&&\{\nabla^k_-,\nabla^l_-\}=0~,\q \{\bar{\nabla}_{k-},\bar{\nabla}_{l-}\}
=0~,
\q \{\nabla^k_-,\bar{\nabla}_{l-}\}=2i\delta^k_l\nabla_=~,\lb{a11}
\eea
where $\nabla$ are the covariant derivatives in the
$(4|8,8)$-dimensional superspace, $W_{kl}$ and $W^{kl}$ are the
gauge-covariant superfields constructed from the gauge connections. These
superfields satisfy the  subsidiary conditions
\be
W^{ik}\equiv \overline{W_{ik}}=-{1\over2}\varepsilon^{ikjl}W_{jl}~.
\lb{A11b}
\ee

The equations of motion for the superfield strengthes follow from the
Bianchi identities
\bea
&& \nabla^i_\pm W^{kl} + \nabla^k_\pm W^{il}=0~,\nn\\
&& \bar{\nabla}_{i\pm} W^{kl}={1\over2}(\delta^k_i
\bar{\nabla}_{j\pm}W^{jl}- \delta^l_i\bar{\nabla}_{j\pm}
W^{jk})~.\lb{a12}
\eea

Let us consider the light-cone gauge conditions
\be
A^k_+=0~,\q \bar{A}_{k+}=0~,\q A_\pp =0~.\lb{a14}
\ee

We shall use the $SU(4)/ U(1)^3$ harmonics \cite{Ba,AFSZ,IKNO} for the HSS
interpretation of the non-Abelian $N=4$ constraints (\ref{a8}-\ref{a11})
by analogy with the Abelian case.

The $SU(4)/U(1)^3$ harmonics parametrize the
corresponding coset space. They form an $SU(4)$ matrix and are
defined modulo $U(1)\times U(1)\times U(1)$ transformations
\be
u^1_i=u^{(1,0,1)}_i~,\quad u^2_i=u^{(-1,0,1)}_i~,\quad
u^3_i=u^{(0,1,-1)}_i~,\quad u^4_i=u^{(0,-1,-1)}_i~\lb{I1} \ee
where $i$ is the index of the quartet representation of $SU(4)$.
The complex conjugated harmonics have opposite $U(1)$ charges
\be
u_1^i=u^{i(-1,0,-1)}~,\quad u_2^i=u^{i(1,0,-1)}~,\quad
u_3^i=u^{i(0,-1,1)}~,\quad u_4^i=u^{i(0,1,1)}~.\lb{I2} \ee
Note
that we use indices $I, J=1, 2, 3, 4$ for the projected components
of the harmonic matrix which do not transform with respect to the
'ordinary' $SU(4)$ transformations. The authors of Ref.\cite{HH} prefer
to use the $SU(4)/S(U(2)\times U(2))$ harmonics for the $N=4$ theory.

The corresponding harmonic derivatives $\partial^I_J$ act on these
harmonics and satisfy the $SU(4)$ algebra.

The special  conjugation of the $SU(4)$ harmonics has the following form:
\be
u^1_i\leftrightarrow u^i_4~,\q u^2_i\leftrightarrow u^i_3~,\q
u^3_i\leftrightarrow u^i_2~,\q u^4_i\leftrightarrow u^i_1\lb{I14}
\ee
and the conjugation of the harmonic derivatives is
\be
\partial^1_2 f\leftrightarrow-\partial^3_4\widetilde{f},\q
\partial^1_4 f\leftrightarrow-\partial^1_4\widetilde{f},
\ee
where $f(u)$ is an arbitrary harmonic function.

The analytic coordinates in the $N=4$ superspace $H(4,12|6,6)$
are
\bea
&&\zeta=(X^\pp,~X^=,~Y,~\bar{Y}~| \theta^\pm_2 ,~
\theta^\pm_3,~\theta^\pm_4,~\bar\theta^{1\pm},~\bar\theta^{2\pm},~
\bar\theta^{3\pm})\;,\q
X^\pp=\xp +i(\tpf\btpf -\tpo\btpo)\;,\nn\\
&&X^= =\xm +
i(\tmf\btmf -\tmo\btmo)\;,\q Y=y+i(\tpf\btmf -\tpo\btmo)\;,\nn\\
&&\bar{Y}=\bar{y}+i(\tmf\btpf -\tmo\btpo)\;,\q
\theta^\pm_I=\theta^\pm_k u^k_I~,\q\bar\theta^{I\pm}=
\bar\theta^{\pm k}u_k^I~.
\lb{I15}
\eea

The spinor derivatives have the following simple form in these
coordinates:
\bea
&&D^1_\pm =\partial^1_\pm~,\qq
\bar{D}_{4\pm} =\bar\partial_{4\pm}~,  \lb{I16} \\
&& D^2_+ =\partial^2_+ +i\bar\theta^{2+}\Pp+i\bar\theta^{2-}\PY~,
\q  D^2_- =\partial^2_- +i\bar\theta^{2+}\bPY+i\bar\theta^{2-}\Pm~,\\
&&
\bar{D}_{1+} =\bar\partial_{1+} +2i\theta^+_1\Pp+2i\theta^-_1\bPY~,
\q \bar{D}_{1-} =\bar\partial_{1-} +2i\theta^+_1\PY+2i\theta^-_1\Pm~.
 \eea

The corresponding harmonic derivatives are
\bea
 &&\Dot =\partial^1_2
+i\tpt\btpo\Pp+i\tpt\btmo\PY+i\tmt\btpo\bPY
+i\tmt\btmo\Pm\nn\\
&&-\tpt\ppo-\tmt\pmo+\btpo\bppt+\btmo\bpmt
~,\\
 &&\Dhf =\partial^3_4
+i\tpf\btph\Pp+i\tpf\btmh\PY+i\tmf\btph\bPY
+i\tmf\btmh\Pm\nn\\
&&-\tpf\pph-\tmf\pmh+\btph\bppf+\btmh\bpmf
~.
\eea
Other projections of the  Grassmann and harmonic derivatives can be
constructed analogously.

Let us consider the harmonic projections of the CB covariant derivatives
and the corresponding  connections
\bea
&&\nabla^I_+=u^I_k \nabla^k_+=D^I_+~,\qq \bN_{I+}=u^j_I\bN_{j+}
=\bar{D}_{I+}~,\\
&&\nabla^I_-=u^I_k \nabla^k_-=D^I_- +\cA^I_-~,\qq \bN_{I-}=u^j_I\bN_{j-}
=\bar{D}_{I-}+\bcA_{I-}~.
\lb{B7}
\eea

Taking into account these relations we can transform
the CB-constraints (\ref{a8}-\ref{a11}) to the equivalent
(2,2)-dimensional set of the G-integrability relations:
\be
\{\N^1_\pm,\N^1_\pm\}=\{\N^1_\pm,\bN_{4\pm}\}=\{\bN_{4\pm},
\bN_{4\pm}\}=0~.\lb{B8}
\lb{B10}
\ee

Thus, the $N=4$ SYM-geometry preserves the Grassmann (6,6) analyticity.
It can be shown that the covariant (4,4)-analyticity of superfield
strength $u^1_ku^2_kW^{ik}$ follows from the basic (6,6)-analyticity in
the HSS geometric formalism.

Now we shall discuss the solution of the G-integrability
relations
\be
 \cA^1_{\pm}(v)=e^{-v}\left(D^1_\pm
e^v\right)~,\q \bcA_{4\pm}(v)=e^{-v}
\left(\bar{D}_{4\pm}e^v\right)~,\lb{B15}
\ee
where $v(z,u)$ is the superfield bridge matrix.

The gauge transformations of the bridge
\be
e^v~\Rightarrow~e^\lambda e^ve^{-\tau}~,\lb{lambda}
\ee
contain the (6,6)-analytic  AB-gauge  parameters $\lambda$
\be
(D^1_\pm, \bar{D}_{4\pm}) \lambda =0\lb{B16}
\ee
and   the harmonic-independent constrained
CB-gauge parameters $\tau$.

Matrix $e^v$ determines a transform of the CB-gauge superfields
to the analytic basis (AB).
The analytic gauge group acts on the harmonic connections in AB
\bea
&&\nabla^I_K=e^vD^I_Ke^{-v}=D^I_K+V^I_K(v)~,\lb{br1}\\
&&\delta V^I_K =D^I_K\lambda +[V^I_K,\lambda]~.
\lb{B16b}
\eea

Our gauge choice $\cA^1_+=\bcA_{4+}=0$ corresponds to the following
partial gauge conditions for the bridge:
\be
(\Dpo, \bDpf) v=0~.\lb{gauge}
\ee

We treat bridge $v$ as the basic on-shell superfield, so the
 SYM-equations of this approach are formulated  for this
superfield
\be
[D^I_K,e^{-v}\Dmo e^v]=[D^I_K,e^{-v}\bDmf e^v]=0~,\q I < K.
\lb{Hanal}
\ee

The subsidiary condition \p{A11b} is equivalent to the reality
condition  for the harmonic projection of the superfield strength
$u^1_iu^2_kW^{ik}$ \cite{Ba,HH} and corresponds to the following equation
in the bridge representation:
\be
-\Dpt (e^{-v}\Dmo e^v)=\bDph(e^{-v}\bDmf e^v)~.\lb{selfd}
\ee

By analogy with the $N=3$ formalism \cite{NZ} one can choose the following
light-cone  gauge for the $N=4$ bridge:
\be
v=\tmo b^1 +\btmf \bar{b}_4 +\tmo\btmf d^1_4~,\lb{gauge2}
\ee
where the fermionic matrices $b^1, \bar{b}_4$ and the bosonic matrix
$d^1_4$ are the (6,6) analytic superfields. This bridge is nilpotent
\bea
&& v^2=\tmo\btmf
[\bar{b}_4,b^1]\;,\q v^3=0~,\lb{nilp1}\\
&&e^{-v}= I - v+{1\over2}v^2=I-\tmo b^1- \btmf \bar{b}_4+\tmo\btmf
({1\over2}[\bar{b}_4,b^1]-d^1_4 )~.\lb{nilp2}
\eea

In the gauge group $SU(n)$, our superfields satisfy the conditions
\be
(b^1)^\dagger=\bar{b}_4~,\q (d^1_4)^\dagger=-d^1_4~,\q \T b^1=\T d^1_4=0~.
\ee

Consider the parametrization of the basic spinor connections
in our gauge
\bea
&&\cA^1_-(v)=b^1-\tmo (b^1)^2+\btmf f^1_4+
\tmo\btmf[b^1,f^1_4]~,\lb{cbcon1}\\
&&\bcA_{4-}(v)=\bar{b}_4-\btmf (\bar{b}_4)^2+\tmo\bar{f}^1_4
-\tmo\btmf
[\bar{b}_4,\bar{f}^1_4]~,\lb{cbcon2}
\eea
where the following auxiliary superfields are introduced:
\be
f^1_4=d^1_4-{1\over2}\{b^1,\bar{b}_4\}\,,\q
\bar{f}^1_4=-d^1_4-{1\over2}\{b^1,\bar{b}_4\}\,.\lb{auxil}
\ee

The H-analyticity equations $(\Dot, \Doh, \Dtf, \Dhf)\cA^1_-(v)=0$
are equivalent to the following (6,6)-analytic equations:
\bea
&& (\Dot, \Doh) b^1=-(\tmt, \tmh) (b^1)^2\;,\q (\Dtf, \Dhf) b^1=-(\btmt,
\btmh)f^1_4\;,
\lb{Coef1}\\
&& (\Dot, \Doh) f^1_4=(\tmt, \tmh)[f^1_4,b^1]~,\q(\Dtf, \Dhf) f^1_4=0~.
\lb{Coef4}
\eea

We shall discuss below the relations between the matrices $b^1$ and
$\bar{b}_4$ which arise from the transform of the CB-condition
\p{selfd} to the analytic representation.

Remember that the following covariant Grassmann derivatives are flat in
the  AB-representation of the gauge group before the gauge fixing:
\be
e^v\nabla^1_\pm e^{-v}=D^1_\pm\;,\q
e^v\bar{\nabla}_{4\pm}e^{-v}=\bar{D}_{4\pm}\;\lb{C1}
\ee

 The harmonic connections in the bridge representations $V^I_K(v)$
\p{br1} satisfy automatically the harmonic
zero-curvature equations
\be
D^I_KV^J_L-D^J_LV^I_K+[V^I_K,V^J_L]=\delta^J_KV^I_L-\delta^I_LV^J_K~.
\ee

Basic SYM-equations \p{Hanal}are equivalent to the dynamical
G-analyticity conditions
\be
(\Dmo, \bDmf)V^I_K(v)=0~,\q I < K~.\lb{Ganal}
\ee

 In  gauge \p{gauge2}, these  equations
give us the following  relations:
\bea
&&V^1_2(v)=\tmt b^1~,\lb{prep1}\q V^1_3(v)=\tmh b^1
~,\lb{prep2}\\
&&V^3_4=(V^1_2)^\dagger=-\btmh \bar{b}_4~,\lb{prep3}\q
V^2_4=(V^1_2)^\dagger=-\btmt \bar{b}_4~,\lb{prep4}
\eea
where all connections are nilpotent.
Similar relations have been considered in the harmonic formalism
of the $N=3$ SYM-theory \cite{NZ}.

 One can also construct the
non-analytic harmonic connections
\be
e^v\Dto e^{-v}=V^2_1=-\tmo \Dto b^1~.
\lb{sol1}
\ee

The conjugated harmonic connection depend, respectively, on  matrix
$\bar{b}_4$ only
\be
V^4_3=(V_1^2)^\dagger=-\btmf \Dfh \bar{b}_4
\;.\lb{sol3}
\ee

It is not difficult to show that the harmonic AB-connections
$V^2_1$  satisfies the partial (8,6)-analyticity condition
\be
\bar{D}_{4\pm} \Vto=0\lb{B23}
\ee
and the conjugated connection possesses the  (6,8)-analyticity
\be
D^1_\pm \Vfh=0~.\lb{B24}
\ee

The basic AB-superfield strengthes can be constructed in terms of the
harmonic connections
\bea
&&w^{12}=-\Dpo\Dmo V^2_1
=-\Dpt b^1
\;,\\
&&w_{34}=-\bDpf\bDmf V^4_3=
-\bDph c_4 \;.
\lb{H7}
\eea
They satisfy the non-Abelian G- and H-analyticity conditions
which generalize the shortness conditions for the corresponding
Abelian superfields \cite{AFSZ}.

The reality condition
\be
w^{12}=-w_{34}  \lb{B27}
\ee
 is equivalent to the single linear differential relation
between the matrices  $b^1$ and $\bar{b}_4$ which  can be easily solved
via the following representation with the anti-Hermitian (6,6)-analytic
bosonic matrix $A^{13}$
\bea
&&b^1=\bDph A^{13}~,\qq \bar{b}_4=\Dpt A^{13}~,\lb{solut}\\
&&w^{12}\equiv -w_{34}=-\Dpt\bDph A^{13}~.
\eea

Consider the evident relation
\be
(b^1)^2={1\over2}\bDph[A^{13},\bDph A^{13}]~.
\ee

Equations \p{Coef1} generate the following relations for $A^{13}$
\be
(\Dot, \Doh) A^{13}={1\over2}(\tmt, \tmh) [A^{13},\bDph A^{13}] ~.
\lb{Aeq1}
\ee

Thus, the harmonic-superspace representation and  light-cone
gauge conditions simplify significantly the analysis of the $N=4$
SYM-equations. We hope that this representation allows us to construct
the interesting solutions of these equations.

The author is grateful to J. Niederle, E. Sokatchev and E. Ivanov
for interesting discussions.
This work is  partially supported
by the grants RFBR-99-02-18417,
RFBR-DFG-99-02-04022 and NATO-PST.CLG-974874.

\renewcommand\baselinestretch{0.6}

\end{document}